\documentclass[a4paper,12pt]{article}
\usepackage{citeref}

\input cohmacros.sty

\def\at#1{[*** \att #1 ***]}  
\def\at#1{} 

\advance\textheight2.7cm        
\advance\topmargin-2.0cm
\advance\textwidth2.6cm
\advance\evensidemargin-1.6cm
\advance\oddsidemargin-1.6cm

\begin{document}

\vspace*{-2cm}
\begin{center}
{\LARGE \bf Foundations of quantum physics} \\[4mm]

{\LARGE \bf  IV. More on the thermal interpretation} \\

\vspace{1cm}

\centerline{\sl {\large \bf Arnold Neumaier}}

\vspace{0.5cm}

\centerline{\sl Fakult\"at f\"ur Mathematik, Universit\"at Wien}
\centerline{\sl Oskar-Morgenstern-Platz 1, A-1090 Wien, Austria}
\centerline{\sl email: Arnold.Neumaier@univie.ac.at}
\centerline{\sl \url{http://www.mat.univie.ac.at/~neum}}

\end{center}


\hfill May 19, 2019

\vspace{0.5cm}

\bigskip
\bfi{Abstract.}
This paper continues the discussion of the thermal interpretation of
quantum physics. 
While Part II and Part III of this series of papers explained and 
justified the reasons for the departure from tradition, the present 
Part IV summarizes the main features and adds intuitive explanations 
and new technical developments. 

It is shown how the spectral features of quantum systems and an 
approximate classical dynamics arise under appropriate conditions.

Evidence is given for how, in the thermal interpretation, 
the measurement of a qubit by a pointer q-expectation may result in a 
binary detection event with probabilities given by the diagonal entries 
of the reduced density matrix of the prepared qubit. 

Differences in the conventions about measurement errors in the thermal 
interpretation and in traditional interpretations are discussed in 
detail.

Several standard experiments, the double slit, Stern--Gerlach, and 
particle decay are described from the perspective of the thermal 
interpretation.

\vfill
For the discussion of questions related to this paper, please use
the discussion forum \\
\url{https://www.physicsoverflow.org}.

\newpage
\vspace*{-2.5cm}
\tableofcontents 

\vspace{2cm}

\newpage
\section{Introduction}

This paper, the fourth of the series on the foundations of quantum 
physics, continues the discussion of the thermal interpretation of
quantum physics by summarizing the main features, by adding intuitive 
explanations and by introducing new technical developments. 

In this introductory section, we first show that the notion of 
indistinguishability of the individual constituents of a system 
naturally leads to the thermal interpretation view. 
We then summarize the most important points from the thermal 
interpretation, discussed in detail in Part II \cite{Neu.IIfound} and 
applied to measurement in Part III \cite{Neu.IIIfound} of this series 
of papers. 

Section \ref{s.classical} shows how the spectral features of quantum 
systems and an approximate classical dynamics arise under appropriate 
conditions.

In Section \ref{s.singleQubits}, formal evidence is given for how, in 
the thermal interpretation, the measurement of a qubit by a pointer 
q-expectation may result in a binary detection event with probabilities 
given by the diagonal entries of the reduced density matrix of the 
prepared qubit. 

In the thermal interpretation, the true properties of a quantum system, 
approximately revealed by a measurement, are the q-expectations rather 
than the eigenvalues. After nearly a century of conditioning to the 
opposite convention specified in Born's rule, this radical change of 
interpretation seems at first sight very counterintuitive. 
A detailed justification and comparison with the traditional convention 
is given in Section \ref{s.errors} from the measurement point of view.
An analysis of the double slit experiment leads to the picture of a 
quantum bucket for measuring a continuous variable with a device 
capable only of producing discrete results.

The final Section \ref{s.currents} shows how the notion of quantum 
currents may be used to visualize in the thermal interpretation the 
finite time dynamics of particle decay and the Stern--Gerlach 
experiment. In the case of observing angular momentum, the 
measurement process is claimed to systematically introduce $O(\hbar)$ 
perturbations of the same kind as rounding errors in floating-point 
computations -- a tiny amount for all but microscopic measurements. 
Due to the representation theory of the compact rotation group, these 
discretize the response of the measurement device to the continuous 
signal represented by the q-expectation $\<J\>$ of the vector-valued 
angular momentum $J$ of the measured particle. This results (depending 
on the precision of the measuring device) in almost\footnote{
In the past, this experiment (and others) could be used for precision 
measurements of $\hbar$. But from May 20, 2019 onwards, $\hbar$ has by 
convention a fixed (but irrational) value, as part of the 2019 
redefinition of SI base units \cite{SIunits}. From then on, one can get 
(by calibration) exact multiples of $\half\hbar$, as claimed in Born's 
rule. However, the thermal interpretation asserts that, since the
measurement results are not reproducible, this seeming exactness of the
angular momentum measurement is a spurious artifact of measuring it 
with a quantum bucket.
} 
exact multiples of $\half\hbar$, with resulting discretization errors 
of order $O(\hbar)$. For a single electron spin measured in 
a Stern--Gerlach experiment, this perturbation is of the same order as 
the size of each component of $\<J\>$, which is bounded itself by 
$\half\hbar$, resulting -- in the convention of the thermal 
interpretation -- in intrinsic measurement errors of up to 200 percent. 
But there is no logical problem since any number of order $O(\hbar)$ is 
an $O(\hbar)$ perturbations of any other number of order $O(\hbar)$.

\bigskip

In the bibliography, the number(s) after each reference give the page 
number(s) where it is cited.

\bigskip
{\bf Acknowledgments.}
Earlier versions of this paper benefitted from discussions with
Rahel Kn\"opfel.

\subsection{Properties of anonymous collections}

Let $x_k$ ($k=1,\ldots,N$) denote the (real) values of some property of 
a collection of $N$ similar classical objects. If the detailed 
identification 
of the objects is deemed irrelevant for certain purposes, the assignment
of indices to the the individual objects may be dropped, in this way 
anonymizing the data. Indeed, this is a common procedure in the 
statistical practice of handling sensitive data. Once this is done we 
can no longer say which property belongs to which object -- in the 
resulting description, the objects have become anonymous, or 
\bfi{indistinguishable}.

As a consequence, the individual values $x_k$ play no longer a useful 
role in the anonymized collection. From a mathematical point of view, 
only symmetric functions of the $x_k$ retain meaningful information
about the collection. By a well-known theorem, every symmetric 
polynomial (and by taking limits, therefore any symmetric analytic 
function) of the $x_k$ can be written as a function of the power sums
 $\sum x_k^e$ ($e=1,2,3,\ldots$), equivalently, as a function of the
sample expectations $\<x^e\>=N^{-1}\D\sum_{k=1}^N x_k^e$. Some 
discontinuous symmetric functions also play a role, and can be written 
as a function of sample expectations of discontinuous functions. Thus 
all meaningful properties of the anonymized collection are encoded in 
expectations $\<f(x)\>$ of functions of the anonymous value $x$ of
an anonymous object of the collection. These expectations may therefore 
be regarded as the \bfi{beables} of the anonymous classical collection. 

It is precisely this situation that probabiliy theory and statistics 
cater for -- the description of anonymous events, not that of actual 
events! We assign probabilities to anonymous events such as ''casting a 
die gives a six'' (where the indefinite article indicates an anonymous 
die), not to the number of eyes shown on a particular die cast at a 
particular time (which is not a random variable but a fixed, though 
possibly unknown vaue). We estimate the expected lifetime of 
''a 45 year old French male'', not that of Francois Renon from Calais, 
say. And so on. Formally, from what is mathematically modeled,
anonymous objects (whose only properties are expectation values and 
probabilities) are very different from \bfi{typical} objects, which 
are identifiable examples of particular objects (whose properties are 
individual values within observable typical ranges). 

Even before the advent of quantum mechanics, it turned out that, in
classical statistical mechanics, atoms are indistinguishable not only 
due to pracical limitations but in principle, and that there is no 
theoretically conceivable way to distinguish them as individuals -- if 
it were possible, the resulting predictions would have an additional 
entropy of mixing, which is in conflict with the observed
thermodynamical properties of bulk systems. This means that
there are fundamental constraints that forbid the atoms in a classical 
multiparticle system to have individual properties. Thus, in a classical
multiparticle system, the atoms are anonymous objects without an 
identity, and the expectations are the only classical beables. This 
directly leads to the main innovation of the thermal interpretation --
that expectations are beables.

This situation persists in the quantum case, where atoms and elementary 
particles are in principle\footnote{
Exceptions are cases where the range of some quantity identifies a 
unique particle. (This is analogous to the identifiability of outliers 
in anonymous statistical data.) Examples include a single atom prepared 
in an ion trap, single atoms on the surface of some other material, 
the atom closest to a given lattice position in a piece of metal, or, 
in the Hartree-Fock approximation, the outermost electron of an atom. 
In this case, the identification can be made by expectations of 
quantities containing a characteristic function of the defining 
property as a factor.
} 
indistinguishable, too.
Thus the atoms and elementary particles in a multiparticle quantum 
system are also anonymous objects without an identity. This is 
reflected in the fact that on the physical Hilbert space of correctly 
symmetrized wave functions, no particle position operator is definable;
particle positions are spurious objects. The definable operators are
cumulative $N$-particle operators. When expressed in terms of the 
second quantization formalism, these become quantum fields. Thus the
q-expectations of quantum fields are the natural generalizations of the 
classical beables.

\subsection{Thermal interpretation summary}\label{ss.TIsummary}

The thermal interpretation treats quantum physics in a deterministic, 
almost classical fashion. It differentiates between 

\pt
\bfi{properties} of quantum systems -- beables, that the systems 
possess according to a scientific model, independent of whether these 
properties are known or even knowable), and 

\pt 
\bfi{experiments} consisting of a sequence of \bfi{measurements} -- 
which are the scientists's approximate way of checking such properties 
and validating the corresponding models.

In classical mechanics, particles exist. States define their properties 
(the beables of classical mechanics), which are given by the exact 
positions and momenta of the particles, some of which can be 
approximately measured. From a fundamental point of view, fields are 
(as in classical continuum mechanics) only coarse-grained approximate 
concepts. This is the standard interpretation of classical physics.

In quantum field theory, fields exist. States define their properties 
(the beables of quantum field theory), which are given by the exact 
q-expectations of the fields and their appropriately normally ordered 
or time-ordered products, some of which can be approximately measured.  
From a fundamental point of view, particles are (as in classical 
geometric optics) only coarse-grained approximate concepts. This is the 
thermal interpretation of quantum physics. 

Experimental physics is in both cases about how to do the 
measurements, and under which conditions which measurements are how 
accurate. This is achieved using the standard theory based upon three
ingredients: the formal core of quantum mechanics, the respective 
foundations, and

{\bf (CC)} {\bf Callen's criterion} (cf. \sca{Callen} \cite[p.15]{Cal}):
Operationally, a system is in a given state if its properties are 
consistently described by the theory for this state.

The thermal interpretation makes quantum physics as deterministic as 
classical physics, and explains all random quantum effects as resulting 
from coarse-graining, as in classical physics. Quantized measurement 
results (as observed angular momentum measurements) are explained by 
environment-induced randomness and environment-induced dissipation, as 
for a classical, environment-induced diffusion process in a double-well 
potential. Born's statistical interpretation follows, in the limited 
range where it applies, from this and the deterministic rules.

The deterministic Ehrenfest dynamics of the collection of all 
q-expectations couples local q-expectations (e.g., idealized pointer 
readings) to multilocal q-expectations, and accounts in this way for 
the nonclassical correlations observed in long-distance entanglement. 

A subsystem of a composite system is selected by picking a vector space 
of quantities (linear operators) relevant to the subsystem. The 
existence of multilocal q-expectations implies that a composite 
system is more than its parts. 
Regarding a tensor product of two systems as two separate subsystems 
(as often done informally) is appropriate only when all quantities that 
correlate the two systems are deemed irrelevant. If this is not the 
case, thinking of the composite system only in terms of its subsystems 
produces the weird features visible in many discussions of quantum 
entanglement.

\subsection{Advantages of the thermal interpretation}

\nopagebreak
\hfill\parbox[t]{10.8cm}{\footnotesize

{\em [...] auf die objektive Beschreibbarkeit der individuellen 
{\em Makro-}Systeme (Beschreibung des 'Realzustandes') nicht verzichtet 
werden kann ohne dass das physikalische Weltbild gewissermassen sich in 
einen Nebel aufl\"ost. 
Schliesslich ist die Auffassung wohl unvermeidbar, dass die Physik 
nach einer Realbeschreibung des Einzel-Systems streben muss. Die Natur 
als Ganzes kann eben nur als individuelles (einmalig existierendes) 
System gedacht werden und nicht als eine 'System-Gesamtheit'.
}


\hfill Albert Einstein, 1953 \cite[p.40]{Ein53}
}
\bigskip

The conventions embodied in the thermal interpretation have, compared 
to tradition, several direct or indirect advantages. 

\pt
The thermal interpretation allows a consistent quantum description of 
the universe and its subsystems, from the smallest to the largest 
levels of modeling, including its classical aspects.

\pt
The thermal interpretation preserves the agreement of quantum theory 
with the experimental record.

\pt
At the levels of the postulates, the thermal interpretation requires 
much less technical mathematics (no spectral theorem, no notion of 
eigenvalue, no probability theory).

\pt
The foundations are easily stated and motivated since they are
essentially the foundations used everywhere for uncertainty
quantification.

\pt
The thermal interpretation allows one to make definite statements 
about each single quantum system, no matter how large or small it is.

\pt 
The thermal interpretation eliminates from the foundations the 
philosophically problematic notions of probability and measurement.

As a result of the multi-valuedness of the true values, Born's 
statistical interpretation needs probabilities in the very foundations 
of quantum physics. In contrast, in the thermal interpretation, 
probabilities are absent in the foundations of quantum physics, as a 
result of the single-valuedness of the true values.

Every observable quantity $A$ has an associated intrinsic 
state-dependent uncertainty $\sigma_A$ within which it can be determined
(in principle). According to the thermal interpretation it is as
meaningless to ask for more accuracy as to ask for the position of 
an apple to mm accuracy. Statistics enters whenever a single value has 
too much uncertainty, and only then. In this case, the uncertainty can 
be reduced by calculating means, as within classical physics. 

\pt
Position and momentum of distinguishable particles have at any time
simultaneously idealized but uncertain values (just like the position 
and momentum of a classical rocket), eliminating the spooky nature of 
the traditional quantum ontologies.

\pt
The thermal interpretation solves the measurement problem, 
makes quantum mechanics much less mysterious, and makes it much less 
different from classical mechanics. 

That quantities with large relative uncertainty (such as single spins) 
are erratic in measurement is nothing special to quantum physics but 
very familiar from the measurement of classical noisy systems. 
The thermal interpretation asserts (and gives good grounds for trusting)
that all uncertainty is of this kind, and probabilities enter only at 
the same level as in classical physics -- as residual uncertainty of 
approximate, coarse-grained treatments. 
 
\pt
Open problems concerning technical details (mentioned at the end of 
Part III) provide mathematical challenges, and show that, unlike 
traditional interpretations, the thermal interpretation is in principle 
refutable by theoretical arguments.

\section{Classical and spectral features of quantum physics}
\label{s.classical}

\nopagebreak
\hfill\parbox[t]{10.8cm}{\footnotesize

{\em Es ist w\"unschenswert, die folgende Frage m\"oglichst elementar 
beantworten zu k\"onnen: Welcher R\"uckblick ergibt sich vom Standpunkt
der Quantenmechanik auf die Newtonschen Grundgleichungen der 
klassischen Mechanik? 
}

\hfill Paul Ehrenfest 1927 \cite{Ehr}
}
\bigskip

In 1927, when the Copenhagen interpretation (the informal agreement on 
the interpretation reached at the 1927 Como and Solvay conferences) was 
forged, its main purpose was to reconcile the then new quantum 
formalism with the experimental evidence available at that time. Apart 
from the Stern--Gerlach experiment, the evidence consisted exclusively 
of (i) the observation of spectra of atoms and molecules, and (ii) the 
need to reconcile the quantum description of the invisible microscopic 
details with the classical description of the macroscopic world.

This section shows that the same evidence is naturally explained by 
the thermal interpretation. Indeed, with a little more work and 
imagination, Paul \sca{Ehrenfest}, whose paper \cite{Ehr} appeared in 
1927, could have easily found and justified this interpretation.

\subsection{The classical approximation}

We consider an interacting multiparticle quantum system with mass 
matrix $M$, position operator $q$, and momentum operator $p$, with 
dynamics given by the Hamiltonian $H=\half p^TM^{-1}p+V(q)$.
To arrive at an approximate classical equation of motion for the 
q-expectation $\ol q=\<q\>$, we apply the Ehrenfest equation and find,
using the canonical commutation relations and componentwise 
expectations, the formulas
\[
\frac{d}{dt}\<q\>=M^{-1}\<p\>,~~~\frac{d}{dt}\<p\>=-\<\nabla V(q)\>,
\]
hence the equation
\lbeq{e.EhrTrad}
M\frac{d^2}{dt^2}\ol q+\<\nabla V(q)\>=0.
\eeq
by \sca{Ehrenfest} \cite{Ehr}, who observed the close formal 
relationship with the classical equation of motion 
\lbeq{e.clDyn}
M\frac{d^2}{dt^2}q+\nabla V(q)=0
\eeq
for this Hamiltonian. To turn this formal relationship into a 
quantitative approximation, we first prove the following 

{\bf Approximation Lemma.} Let $f$ be a twice continuously 
differentiable complex-valued function on $\Rz^n$. Then, for every
vector $q$ of $n$ commuting self-adjoint quantities with convex joint 
spectrum and every state, we have (with the spectral norm)
\lbeq{e.apLemma}
|f(q)-f(\ol q)|\le\half\|f''(q)\|\sum_{k=1}^n \sigma_{q_k}^2.
\eeq
Indeed, for any $\wt q$ in the joint spectrum of $q$ and 
$\eps=\wt q-\ol q$, we have 
\[
f(\wt q)=f(\ol q+\eps)=f(\ol q)+f'(\ol q)\eps
     +\int_0^1 \eps^Tf''(\ol q + s\eps)\eps sds.
\]
By assumption, $\ol q + s\eps$ is for all $s\in[0,1]$ in the joint 
spectrum of $q$, hence by definition of the spectral norm,
\[
\Big|\frac{\eps^Tf''(\ol q + s\eps)\eps}{\eps^T}\eps\Big|
\le \|f''(\ol q + s\eps)\|_2\le \|f''(q)\|. 
\]
Therefore 
\[
\bary{lll}
\Big|f(\wt q)-f(\ol q)-f'(\ol q)(\wt q-\ol q)\Big|
&\le& \D\int_0^1 \|f''(q)\|\eps^T\eps sds\\
&=&\half\|f''(q)\|\eps^T\eps
=\half\|f''(q)\|\D\sum_{k=1}^n (\wt q_k-\ol q_k)^2,
\eary
\]
This inequality therefore also holds for $q$ in place of $\wt q$. 
Taking q-expectations, we find 
\[
\bary{lll}
|\<f(q)-f(\ol q)\>|
&=&\Big|\Big\<f(q)-f(\ol q)-f'(\ol q)(q-\ol q)\Big\>\Big| \\[2mm]
&\le& \half\|f''(q)\|\D\sum_{k=1}^n \Big\<(q_k-\ol q_k)^2\Big\>
=\half\|f''(q)\|\D\sum_{k=1}^n \sigma_{q_k}^2,
\eary
\]
proving the lemma. Returning to our original goal, we rewrite 
\gzit{e.EhrTrad} in the form
\[
M\frac{d^2}{dt^2}\ol q+\nabla V(\ol q)=-\<\nabla V(q)-\nabla V(\ol q)\>
\]
and apply the approximation lemma to the right hand side. Under the 
assumption that the potential $V$ is three times continuously 
differentiable and the spectrum of the third derivative $V'''(q)$ is 
bounded by a constant $C$, we may conclude the differential inequality
\[
\Big|M\frac{d^2}{dt^2}\ol q+\nabla V(\ol q)\Big|
\le C\sum_{k=1}^n \sigma_{q_k}^2.
\]
Thus as long as the uncertainties $\sigma_{q_k}$ remain sufficiently 
small, the classical dynamical law \gzit{e.clDyn} holds with good 
accuracy for the q-expectation $\ol q$ in place of $q$. 

Under these conditions, which hold by the weak law of large numbers 
whenever $q$ refers to the center of mass coordinates of macroscopic 
spherical bodies at macroscopic distances from each other, the 
q-expectations satisfy the traditional classical equation of motion.
This proves that Newton's mechanics is a macroscopic approximation to 
the quantum dynamics of q-expectations. 

Thus classical physics emerges without any difficulty from the thermal 
interpretation together with the weak law of large numbers.

In a similar way, one can justify the quantum classical dynamical 
models discussed in Subsection 4.6 of Part III \cite{Neu.IIIfound}, 
which treat only some low uncertainty quantities as classical and 
keep the quantum nature of the remaining ones.

\subsection{The Rydberg--Ritz combination principle}

Here we show that in any quantum system, the differences of the energy 
levels (the eigenvalues of the Hamiltonian $H$) are in principle 
directly observable, since they represent excitable oscillation 
frequencies of the system and thus can be probed by coupling the system 
to a harmonic oscillator with adjustable frequency. Thus the observed 
spectral properties of quantum systems appear in the thermal 
interpretation as natural resonance phenomena.

To see this, we shall assume for simplicity a quantum system whose 
Hamiltonian has a purely discrete spectrum. For a partially continuous 
spectrum, analogous results, in which sums are replaced by Stieltjes 
integrals, can be proved using the Gel'fand--Maurin theorem, also known 
under the name nuclear spectral theorem (cf. \sca{Maurin} \cite{Mau}).

We work in the Heisenberg picture in a basis of eigenstates of the 
Hamiltonian, such that $H|k\>=E_k|k\>$ for certain energy levels $E_k$.
The q-expectation 
\[
\<A(t)\>=\Tr \rho A(t)=\sum_{j,k}\rho_{jk}A_{kj}(t)
\]
is a linear combination of the matrix elements 
\[
A_{kj}(t)=\<k|A(t)|j\>=\<k|e^{iHt/\hbar}Ae^{-iHt/\hbar}|j\>
=e^{iE_kt/\hbar}\<k|A|j\>e^{-iE_jt/\hbar}
=e^{i\omega_{kj}t}\<k|A|j\>,
\]
where 
\lbeq{e.RR}
\omega_{kj} = \frac{E_k-E_j}{\hbar}.
\eeq
Thus the q-expectation exhibits multiply periodic oscillatory behavior
whose frequencies $\omega_{jk}$ are scaled differences of energy levels.
This relation, the modern form of the \bfi{Rydberg--Ritz combination 
principle} found in 1908 by \sca{Ritz} \cite{Ritz}, may be expressed 
in Planck's form\footnote{
The formula \gzit{planck-E} appears first in the famous 1900 paper by
\sca{Planck} \cite{Pla} on the radiation spectrum of a black body. 
Planck wrote it in the form $\Delta E = h\nu$, where $h=2\pi\hbar$ and 
$\nu=\omega/2\pi$ is the linear frequency. The symbol for the quotient 
$\hbar=h/2\pi$, which translates this into our formula was invented 
much later, in the 1930 quantum mechanics book by \sca{Dirac} 
\cite{Dir}.
} 
\lbeq{planck-E}
\Delta E = \hbar\omega.
\eeq
To probe the spectrum of a quantum system, we bring it into contact 
with a macroscopically observable (hence classically modeled) weakly 
damped harmonic oscillator. For simplicity we treat just a single 
harmonic oscillator. In practice, one often observes many
oscillators simultaneously, e.g., by observing the oscillations of the
electromagnetic field in the form of electromagnetic radiation --
light, X-rays,  or microwaves. However, in most cases the oscillators 
may be regarded as independent and noninteracting. The result of
probing a system with multiple oscillators results in a linear
superposition of the results of probing with a single oscillator.
This is a special case of the general fact that solutions of linear
differential equations depend linearly on the right hand side.

From the point of view of the macroscopically observable classical 
oscillator, the probed quantum system appears simply as a time-dependent
external force $F(t)$ that modifies the dynamics of the free
harmonic oscillator. Instead of the harmonic equation 
$m \ddot q + c \dot q + k q =0$ with real $m,c,k>0$, we get the 
differential equation describing the \bfi{forced harmonic oscillator}, 
given by
\[
m\ddot q + c\dot q + kq =F(t),
\]
where the external force $F$ is the q-expectation
\[
F(t)=\<A(t)\>
\]
of a quantity $A$ from the probed system. We assume the oscillator to 
have an adjustable frequency
\[
\omega=\sqrt{\frac{k}{m}}>0
\]
and consider the response as a function of $\omega$ at fixed mass $m$ 
and stiffness $k=m^2\omega$. 

If the measurement is done far from the probed system, such as a 
measurement of light (electromagnetic radiation) emitted by a 
far away source (e.g., a star, but also a Bunsen flame observed by 
the eye), the back reaction of the classical oscillator on the probed 
system can be neglected. Then the probed system can be considered as 
isolated and evolves according to the preceding analysis, hence the 
external force $F$ can be written as a superposition
\[
F(t) = \sum_l F_l e^{i\omega_l t},
\]
of exponentials oscillating with the (positive and negative)
Rydberg--Ritz frequencies, rearranged in linear order. The solution to 
the differential equation consists of a particular solution and a 
solution to the homogeneous equation. Due to damping, the latter is 
transient and decays to zero. There is a distinguished particular 
solution persisting after the transient decayed, which oscillates with 
the same frequencies as the force, easily seen to be given by
\[
q(t) = \sum_l q_l e^{i\omega_l t},~~~~
q_l = \frac{F_l}{m(\omega^2-\omega_{l}^{2}) + ic\omega_l}.
\]
Since the frequencies are real and distinct, the denominator cannot 
vanish. The energy in the $l${th} mode is therefore proportional to
the amplitude 
\lbeq{e.lorentzshape}
|q_l|^2 
=\frac{|F_l|^2}{m^2(\omega^2k-\omega_{l}^{2})^2 + c^2\omega_l^2},
\eeq
with a maximum at the resonance frequency $\omega=|\omega_l|$.
The total energy is proportional to 
\lbeq{e.q2}
|q(t)|^2=\sum_l |q_l|^2+\sum_{k\ne l}q_k^*q_le^{i(\omega_k-\omega_l)t}.
\eeq
We now look at the short-time average (recorded by a typical detector).
If the frequencies $\omega_k$ with significant intensity are 
well-separated, the oscillating terms in \gzit{e.q2} cancel out 
and we find a total mean energy proportional to
\[
a(t)\approx \sum_l |q_l|^2
=\sum_l \frac{|F_l|^2}{(m^2(\omega^2k-\omega_{l}^{2})^2+c^2\omega_l^2}.
\]
As a function of the varying frequency, this has the typical 
spectral intensity form of a superposition of Lorentz shaped resonance 
curves, with local maxima very close to the resonance frequencies 
$|\omega_l|$.

\section{Measuring single qubits}\label{s.singleQubits}

In this section we consider in detail how the thermal interpretation
explains the emergence of binary responses of a measurement device 
when coubled with the simplest quantum object, a qubit, with 
probabilities given by the diagonal entries of the reduced density 
matrix of the prepared qubit.

\subsection{Physical systems and their states}

From a fundamental point of view, each physical system is a subsystem 
of the whole \bfi{universe}, the only truly isolated physical system 
containing the solar system. 

In the standard Schr\"odinger picture, the universe has at each time $t$
a universal density operator 
\[
\rho(t)=e^{-itH/\hbar}\rho(0)e^{itH/\hbar}, 
\]
in terms of which the q-expectations $\<A\>_t=\Tr\rho(t)A$ at time $t$
are defined. In the covariant Schr\"odinger picture introduced in  
Part II \cite{Neu.IIfound}, the universe has at each spacetime position 
$x$ a universal density operator 
\[
\rho(x)=e^{-ip\cdot x/\hbar}\rho(0)e^{ip\cdot x/\hbar}, 
\]
in terms of which the q-expectations $\<A\>_x=\Tr\rho(x)A$ at spacetime 
position $x$ are defined. 
 
A \bfi{physical system} is a subsystem of the universe. It is selected 
by distinguishing the elements of a vector space $\Ez$ of quantities 
(linear operators on the Hilbert space of the universe) as being the 
quantities relevant to the subsystem, and restricting the q-expectation 
mapping of the universe to $\Ez$.

In many cases, the physical system $S$ is defined by a decomposition of 
the Hilbert space $\Hz$ of the universe into a tensor product 
$\Hz=\Hz^S\tensor\Hz^E$ of a \bfi{system Hilbert space} $\Hz^S$ 
and an \bfi{environment Hilbert space} $\Hz^E$ for the remaining part 
of the universe. We call such physical systems \bfi{standard}.

Each standard physical system $S$ has a corresponding reduced density 
operator, given in the standard Schr\"odinger picture by 
$\rho^S(t):=\Tr_E\rho(t)$ and in the covariant Schr\"odinger picture by 
$\rho^S(x):=\Tr_E\rho(x)$, where $\Tr_E$ denotes the partial trace over 
the environment. We call the reduced density operators $\rho^S(t)$ and  
$\rho^S(x)$ the \bfi{state} of the physical system at time $t$ or at 
spacetime position $x$, respectively.
These are {\em the only} states the thermal interpretation is concerned 
with at all -- because these are the states containing precisely the 
information about the q-expectations of operators of the universe 
attached to the system $S$. Indeed, the reduced density operator is 
defined such that for linear operators $A$ on $\Hz^S$ describing system
properties, the q-expectations are given by 
\[
\<A\>_t:=\<A\tensor 1\>_t=\Tr\rho(t)(A\tensor 1)=\Tr\rho^S(t)A,
\]
\[
\<A\>_x:=\<A\tensor 1\>_x=\Tr\rho(x)(A\tensor 1)=\Tr\rho^S(x)A,
\]
\def\prep{\fns{prep}}
where $1$ denotes the identity operator on $\Hz^E$.
Each $\rho^S(t)$ and $\rho^S(x)$ is a Hermitian positive semidefinite 
linear operator on $\Hz^S$ with trace 1. Given any Hermitian positive 
semidefinite linear operator $\rho^S$ on $\Hz^S$ with trace 1, it may 
be possible, by utilizing the laws of Nature and the control facilities 
these impart on humans or machines, to ensure that at some time 
$t_\prep$ (or some spacetime position $x_\prep$), $\rho^S(t_\prep)$ 
resp. $\rho^S(x_\prep)$ approximates $\rho^S$ sufficiently well that 
predictions with $\rho^S$ in place of $\rho^S(t_\prep)$ or 
$\rho^S(x_\prep)$ match experimental checks. In this case we say that 
at time  $t_\prep$ (or spacetime position $x_\prep$), the system $S$ is 
\bfi{prepared} in the state $\rho^S$. How to do this is part of the 
experimental art of \bfi{preparation}. 

If $\rho^S$ has rank 1 then  $\rho^S=\psi\psi^*$ for some 
\bfi{state vector} $\psi$ of norm one (determined by $\rho^S$ up to a 
phase). In this case we say that the system is prepared in the 
\bfi{pure state} $\psi$. Physicists can prepare a system in a pure 
state only when this system has very few degrees of freedom.

\subsection{A single qubit}

We consider a single qubit as a subsystem of the universe. The Hilbert 
space of the universe can be decomposed into a tensor product 
$\Hz=\Hz^S\tensor\Hz^E$ of a 2-dimensional system Hilbert space $\Hz^S$ 
and an environment Hilbert space $\Hz^E$ for the remaining part of the 
universe. We suppose that the qubit is prepared in a state defined by a 
general reduced density matrix $\rho^S$ with components 
$\rho^S_{jk}=\<j|\rho^S|k\>$. Then $\rho^S$ is given by 
\[
\rho^S=\sum_{j,k} \rho^S_{jk}|j\>\<k|.
\]
Since $\rho^S$ is Hermitian positive semidefinite with trace 1,  
\lbeq{e.rhoS}
\rho^S=\pmatrix{p & \alpha^*\cr \alpha & 1-p}
\eeq
for some real number $p\in[0,1]$ and some complex number $\alpha$ with
\lbeq{e.alpha}
|\alpha|\le\sqrt{p(1-p)}.
\eeq
According to the thermal interpretation, the true value of the up 
operator $A=\pmatrix{1 & 0\cr 0 & 0}$ is 
\[
\ol A=\<A\>=\Tr_S\ A = p,
\]
with an uncertainty of 
\[
\sigma_A=\sqrt{\<A^2\>-\ol A^2}=\sqrt{p(1-p)}.
\]
In particular, the true value has no intrinsic uncertainty iff $p=0$ or 
$p=1$.

\subsection{The response of environmental beables} 

In the following we analyze in which way $p$ is reflected in an 
arbitrary environmental q-expectation.
For simplicity, we assume that at preparation time $t=0$, the 
density operator of the universe in the Schr\"odinger picture has the 
tensor product form 
\lbeq{e.rho0}
\rho_0=\rho^S\tensor\rho^E
=\sum_{j,k} \rho^S_{jk}|j\>\<k|\tensor\rho^E.
\eeq
(This assumption could be relaxed but not without going through much 
more technical computations.) The dynamics of the universe is governed 
by a unitary matrix $U(t)$ turning $\rho_0$ into 
\[
\rho(t)=U(t)\rho_0U(t)^*.
\]
We may decompose $U(t)$ uniquely as 
\[
U(t)=\sum_{\ell,k} |\ell\>\<k|\tensor U_{\ell k}(t)
\]
with suitable $ U_{\ell k}(t)\in\Lin\Hz^E$. 

Let $X^E\in\Lin\Hz^E$ be a Hermitian quantity located in the 
environment, so that
\[
X:=1\tensor X^E \in\Lin\Hz
\]
is a quantity of the universe. We want to calculate its q-expectation
\[
\ol X_t:=\<X\>_t=\Tr \rho(t) X=\Tr U(t)\rho_0U(t)^*X
=\Tr \rho_0U(t)^*XU(t)=\Tr \rho_0X(t),
\]
where
\[
\bary{lll}
X(t)&=&U(t)^*XU(t)=U(t)^*(1\tensor X^E)U(t)\\[4mm]
&=&\D\sum_{\ell,j} |j\>\<\ell|\tensor U_{\ell j}(t)^*(1\tensor X^E)
   \D\sum_{\ell',k} |\ell'\>\<k|\tensor U_{\ell' k}(t)\\[6mm]
&=&\D\sum_{\ell,\ell',j,k} |j\>\<\ell|\ell'\>\<k|
                        \tensor U_{\ell j}(t)^*X^EU_{\ell' k}(t)\\[6mm]
&=&\D\sum_{\ell,j,k} |j\>\<k|\tensor U_{\ell j}(t)^*X^EU_{\ell k}(t)
\eary
\]
Using \gzit{e.rho0}, we find that
\[
\bary{lll}
\ol X_t&=&\<X\>_t=\Tr \rho_0X(t)\\[4mm]
&=&\Tr \D\sum_{j',k'} \rho^S_{j'k'}|j'\>\<k'|\tensor\rho^E
 \D\sum_{\ell,j,k}|j\>\<k|\tensor U_{\ell j}(t)^*X^EU_{\ell k}(t)\\[6mm]
&=&\D\sum_{\ell,j,k}\rho^S_{kj}|k\>\<j|
                    \Tr_E\ \rho^EU_{\ell j}(t)^*X^EU_{\ell k}(t).
\eary
\]
If we define $X^S(t)\in\Lin \Hz^S$ by
\lbeq{e.XSt}
X^S(t)_{jk}:=\Tr_E\ \rho^E U_{\ell j}(t)^*X^EU_{\ell k}(t),
\eeq
we arrive at
\[
\ol X_t=\sum_{\ell,j,k}\rho^S_{kj}|k\>\<j|X^S_{jk}(t)
=\Tr_S\ \rho^SX^S(t),               
\]
and by \gzit{e.rhoS},
\lbeq{e.Xt}
\ol X_t=pX^S_{11}(t)+(1-p)X^S_{22}(t)+2\re \alpha X^S_{12}(t).
\eeq

We now consider multiple preparations in the qubit state represented by 
$\rho^S$, but in multiple contexts. We label each such preparation with 
a label $\omega$ from some sample space $\Omega$. Since the split into 
system and environment is different in each preparation, the state 
$\rho^E$ representing the state of the environment depends on the 
preparation label $\omega$. Since $X^E$ was assumed to be Hermitian, 
\gzit{e.XSt} implies that the matrix $X^S(t)$ is also Hermitian and,
being dependent on $\rho^E$, depends on $\omega$. We write
\[
X^S(t)=\pmatrix{\wh x_t(\omega) & \wh z_t(\omega)^* \cr
                \wh z_t(\omega) & \wh y_t(\omega)}
\]
for the realization obtained in the preparation labelled by 
$\omega\in\Omega$. Then we may rewrite \gzit{e.Xt} as
\lbeq{e.Xt2}
\ol X_t
=p\wh x_t(\omega)+(1-p)\wh y_t(\omega)+2\re \alpha\wh z_t(\omega).
\eeq
The $\omega$-dependence is actually a dependence on details of the 
environment that are uncontrollable in practice. Hence it effectively 
turns $X^S(t)$ into a time-dependent random matrix and $\wh x_t$, 
$\wh y_t$ and $\wh z_t$ into time-dependent random variables, of which 
an new realization is obtained for each preparation of the qubit in the 
state represented by $\rho^S$. Their distribution, however, depends on 
more general properties of the environment and is in principle 
amenable to an analysis by the traditional techniques of statistical 
mechanics. Let us write 
\[
X^\eff(t)=\pmatrix{x^\eff_t & {z^\eff_t}^* \cr
                  z^\eff_t & y_t}
\]
for the effective mean of $X^S(t)$, averaged over all preparations 
$\omega\in\Omega$. As a consequence of \gzit{e.Xt}, 
$\ol X_t$ itself behaves like a random variable, with mean 
\[
\ol X^\eff_t=\Tr_S\ \rho^SX^\eff(t)=\<X^\eff(t)\>_S.
\]
Thus we may view every environmental q-expectation as a randomized 
observation of a corresponding effective q-expectation of a quantity 
$X^\eff(t)$ defined on the qubit. Usually, $X^\eff(t)$ is just noise and
$\ol X^\eff_t$ is essentially zero, giving no information about the 
qubit. However, for specially chosen $X$, namely for those where $X$ is 
physically related to the qubit in a significant way, $\ol X^\eff_t$ is 
nonzero and gives nontrivial statistical information about the qubit 
measured -- it is part of a useful measurement device for qubits.

Which qubit quantity is 
observed can be found out by techniques known from quantum tomography.
If $X$ depends on a parameter vector $\theta$, then $X^\eff(t)$ also 
depends on $\theta$, and we can find out the precise $\theta$-dependence
by these techniques. Thus we have an effective way of calibrating our 
measurement device. In particular, whenever we can find a value for 
$\theta$ for which $X^\eff(t)=A$ we get a statistical measurement of 
the true value $p$ of the up operator $A$.

\subsection{The emergence of Born's rule}\label{ss.emergentBorn}

The precise statistical properties of $X^S(t)$ can be found out by 
careful calibration. It can also be predicted by a theoretical analysis 
of the formula defining $X^S(t)$, using the standard techniques of 
statistical mechanics, though this may involve considerable work.
Here we give an outline of how such a theoretical analysis may proceed,
leaving details to future investigations of particular situations
amenable to a more detailed analysis. 

We consider an environmental operator $X^E$ that leads to a 
\bfi{pointer variable}, here a real number $\ol X_t$ that moves in a 
macroscopic time $t>0$ a macroscopic distance to the left (in 
microscopic units, large negative) when $p=0$ and to the right (large 
positive) when $p=1$.
In both cases, $\alpha=0$ by \gzit{e.alpha}, hence by \gzit{e.Xt2}, 
$\ol X_t=\wh x_t(\omega)$ in the first case, and 
$\ol X_t=\wh y_t(\omega)$ in the second case. Therefore
\lbeq{e.gg}
\wh x_t(\omega)\gg 0\gg\wh y_t(\omega).
\eeq
We want to find idealized conditions under which a measurement protocol
produces measurements that follow Born's rule exactly.

In thermodynamics, we get idealized relations in the thermodynamic 
limit of infinite size, which are still applicable with good accuracy
to systems of small but macroscopic size. Similarly, in kinetic theory, 
the scattering matrix, defined through an asymptotic limit of times 
$t\to\pm\infty$ (and the associated infinite separability of clusters) 
is used to define with good accuracy the collision rates and products 
of microscopic scattering events (where distances are small but large 
compared to atomic distances and times are short but large compared to 
the time needed to travel an atomic distance) figuring in the derivation
of the kinetic equations. 

This justifies that we idealize, in the present situation, 
macroscopic distances and times as infinite and therefore assume in 
place of \gzit{e.gg} the exact but idealized limit
\lbeq{e.ideal}
\wh x_t(\omega)\to\infty,~~~\wh y_t(\omega)\to-\infty \for t\to\infty.
\eeq
The desired idealized conditions for the emergence of Born's rule are
now given by the following theorem.

{\bf Theorem.} Let $\wh x_t,\wh y_t,\wh z_t$ be time-dependent random 
variables such that \gzit{e.ideal} holds and
\lbeq{e.x}
\wh v_t(\omega):=\frac{\wh z_t(\omega)}{\wh x_t(\omega)}\to 0 
\for t\to\infty.
\eeq
and define the time-dependent random variable $\wh u_t$ by
\[
\wh u_t(\omega)
:=\frac{\wh x_t(\omega)}{\wh x_t(\omega)-\wh y_t(\omega)}.
\]
If the limiting random variable
\lbeq{e.u}
u:=\lim_{t\to\infty}\wh u_t,
\eeq
exists almost everywhere and is uniformly distributed in $[0,1]$ then
\lbeq{e.prob}
\Pr(\ol X_t\to\infty)=p,~~~\Pr(\ol X_t\to-\infty)=1-p.
\eeq
Indeed, under the stated conditions, $\wh y_t/\wh x_t=1-1/\wh u_t$, 
hence
\[
\ol X_t
=\wh x_t\Big(p+(1-p)\wh y_t/\wh x_t+2\re\alpha \wh z_t/\wh x_t\Big)
=x\Big(1-(1-p)/\wh u_t+2\re\alpha \wh v_t\Big).
\]
Thus
\[
\Pr(\ol X_t\to\infty)=\Pr(1-(1-p)/u>0)=\Pr(u>1-p)=p,
\]
\[
\Pr(\ol X_t\to-\infty)=\Pr(1-(1-p)/u<0)=\Pr(u<1-p)=1-p.
\]
Note that for approximately satisfying Born's rule it suffices that the 
assumptions are satisfied only approximately. Real detectors for
microscopic events often magnify tiny initial displacements in a single
scattering event (a single escaping electron in a photomultiplier or a 
single chemical reaction on a photographic plate) by special processes,
thus making the infinite time limit irrelevant. Moreover, real detectors
have various inefficiencies that may cause deviations from the ideal
probabilistic law expressed by Born's rule.

Assumption \gzit{e.x}
has the nature of a decoherence condition and is likely to be satisfied
under quite general conditions, using a randome phase approximation 
argument. The condition that $u$ exists and is approximately uniformly 
distributed in $[0,1]$ is the essential condition which requires a 
thorough analysis and must be verified in each concrete setting.
It is likely that in the cases treated by AB\&N and B\&P discussed in 
Part III \cite{Neu.IIIfound}, such an analysis can be abstracted from 
their treatment.

\section{Measurement errors}\label{s.errors}

In this section, we give a detailed analysis of the concept of 
meausrement error. This leads to a justification and comparison 
of the convention used to define measurement accuracy in the thermal
 interpretation with the traditional convention. It is followed by an
analysis of the double slit experiment, which exemplifies the crucial 
differences of these conventions.

\subsection{Defining measurement errors}

Measurement errors are ubiquitous in physical practice; their 
definition requires, however, some care. A single measurement produces 
a number, the \bfi{measurement result}. The splitting of the measurement
result into the sum of an intended result (the true value) and a
\bfi{measurement error} (the deviation from it) depends on what one 
declares to be the true value. Thus what can be said about measurement 
errors depends on what one regards as the true value of something 
measured. This true value is a theoretical construct, an idealization 
arrived at by convention. 

Since measured are only actual results, never the hypothesized true 
values, there is no way to determine experimentally which convention is
the right one. Both the quantum formalism and the experimental record 
are independent of what one declares to be the true value of a 
measurement. Different conventions only define different ways of 
bookkeeping, i.e., different ways of splitting the same actual 
measurement results into a sum of true values and errors, in the 
communication about quantum predictions and experiments. Nothing in the 
bookkeeping changes the predictions and the level of their agreement 
with experiment.

Thus the convention specifying what to consider as true values is 
entirely a matter of choice, an \bfi{interpretation}. The convention 
one chooses determines what one ends up with, and each interpretation 
must be judged in terms of its implications for convenience and 
accuracy. Like conventions about defining measurement units 
\cite{SIunits}, interpretations can be adjusted to improvements in 
theoretical and experimental understanding, in order to better serve 
the scientific community.

\bigskip

Born's statistical interpretation of quantum mechanics gives the 
following convention for the prediction of measurement results for 
measuring a quantity given by a self-adjoint operator $A$. One computes 
a number of possible idealized measurement values, the eigenvalues of 
$A$, of which one is exactly (according to most formulations) or 
approximately (if level spacings are below the measurement resolution) 
measured, with probabilities computed from $A$ and the density operator 
$\rho$ by the probability form of Born's rule. Thus the eigenvalues are 
the true values of Born's statistical interpretation. 

Because of the critique of Born's rule given in Part I 
\cite{Neu.Ifound}, the thermal interpretation explicitly rejects the
part of Born's rule that declares the eigenvalues of operators as the 
true values in a measurement. It differs from the tradition created in
1927 by Jordan, Dirac, and von Neumann, and proclaims in direct 
opposition the alternative convention that one computes a single 
possible idealized measurement value, the q-expectation 
\[
\ol A:=\<A\>:=\Tr\rho A
\]
of $A$, which is approximately measured. Thus the true values of the 
thermal interpretation are the q-expectations rather than the 
eigenvalues.

Both interpretations are in full agreement with the experimental record:
The same number obtained by a measurement may be interpreted in a dual 
way: It both measures some random eigenvalue to high (in the 
idealization even infinite) accuracy, and it simultaneously 
measures the q-expectation to low accuracy. 
In both cases, the measurement involves an additional uncertainty 
related to the degree of reproducibility of the measurement, given by 
the standard deviation of the results of repeated measurements.
Tradition and the thermal interpretation agree in that this uncertainty 
is at least 
\[
\sigma_A:=\sqrt{\<A^2\>-\<A\>^2}.
\]
This leads, among others, to Heisenberg's uncertainty relation.

\subsection{What should be the true value?}\label{ss.true}

As an illustration of the differences in the interpretation we first
consider some piece of digital equipment with 3 digit display measuring 
some physical quantity $X$ using $N$ independent measurements. Suppose 
the measurement results were 6.57 in 20\% of the cases and 6.58 in 80\% 
of the cases. Every engineer or physicist would compute the mean 
$\ol X= 6.578$ and the standard deviation $\sigma_X=0.004$ and conclude 
that the true value of the quantity $X$ deviates from $6.578$ by an 
error of the order of $0.004N^{-1/2}$. 

Next we consider the measurement of a Hermitian quantity 
$X\in\Cz^{2\times 2}$ of a 2-state quantum system in the pure up state,
using $N$ independent measurements, and suppose that we obtain exactly 
the same results. The thermal interpretation proceeds as before and 
draws the same conclusion. But Born's statistical interpretation 
proceeds differently and claims that there is no measurement error.
Instead, each measurement result reveals one of the eigenvalues 
$x_1=6.57$ or $x_2=6.58$ in an unpredictable fashion with probabilities 
$p=0.2$ and $1-p=0.8$, up to statistical errors of order $O(N^{-1//2})$.
For $X=\pmatrix{6.578 & 0.004 \cr 0.004 & 6.572}$, both interpretations 
of the results for the 2-state quantum system are consistent with 
theory. However, Born's statistical interpretation deviates radically 
from engineering practice, without any apparent necessity.

Finally we consider the energy measurement of an unknown system with 
discrete, unknown energy levels $E_1<E_2<\ldots$, assumed to be simple 
eigenvalues of the Hamiltonian. We also assume that the system is in a 
pure state $a_1|E_1\>+a_2|E_2\>$, where the kets denote the eigenstates 
of the Hamiltonian and $|a_1|^2=p$, $|a_2|^2=1-p$; for simplicity, 
higher levels than the lowest two are assumed to be absent. As a 
consquence, the q-expectation of the energy (represented by the 
Hamiltonian) can be exactly calculated, giving $\ol E=pE_1+(1-p)E_2$.
The uncertainty of the energy can be exactly calculated, too, giving 
$\sigma_E=\sqrt{p(1-p)}|E_1-E_2|$. 

Something analogous holds for the measurement of any quantity of an
arbitrary 2-state system, such as a spin. According to the experimental
record, the response of a good detector is \bfi{quantized}. Thus the 
measurement results $E$ are concentrated at two spots of the detector, 
just as what one gets when measuring a classical diffusion process in 
a double-well potential (see, e.g., \sca{Hongler \& Zheng} \cite{HonZ}. 
Thus this distribution is bimodal with two sharp peaks, with details 
depending on the detection method used and its resolution.

In a frequently used idealization that ignores the limited efficiency 
of a detector, the distribution may even be assumed to be binomial, with
measurement results that take only one of two values $E_1'$ and $E_2'$
corresponding to the modes of the bimodal distribution. This 
idealization eliminates in particular the effects responsible for a 
detector efficiency of $<100\%$ in real experiments.

According to the thermal interpretation, each measurement result $E$ 
is taken to be an approximation of the true value $\ol E$, with an error
$|E-\bar E|$ of order at least $\sigma_E$. In the limit of arbitrarily
many repetitions, the mean value of the approximations approaches 
$\ol A$ and their standard deviation approaches $\sigma_E$. The bimodal 
distribution of the measurement results is explained by 
environment-induced randomness and environment-induced dissipation, as 
for a classical, environment-induced diffusion process in a double-well 
potential.

According to Born's statistical interpretation in the standard 
formulation,\footnote{
This is the formulation appearing in \sca{Wikipedia} \cite{Wik.Born}. 
\sca{Griffiths \& Schroeter} \cite[p.133]{GriS} declare, ''If you 
measure an observable [...] you are certain to get one of the 
eigenvalues''. \sca{Peres} \cite[p.95]{Peres} defines, ''each one of 
these outcomes corresponds to one of the eigenvalues of $A$; that
eigenvalue is then said to be the result of a measurement of $A$''.
Textbooks such as \sca{Nielsen \& Chuang} \cite[p.84f]{NieC} seem to 
avoid the issue by not referring to eigenvalues at all. But their 
declaration, ''Quantum measurements are described by a 
collection $\{M_m\}$ of measurement operators. [...] The index $m$ 
refers to the measurement outcomes that may occur in the experiment.
[...] the probability that result $m$ occurs'', with a formula that 
summed over all $m$ gives the value $1$, still assumes that the 
values $m$ are exact results -- otherwise each of several approximations
to the same intended result would have to be represented by a different
$M_m$, and their summation would not give $1$. 
} 
''the measured result will be one of the eigenvalues'',
each actual measurement result $E$ is claimed to be one of the the 
exact (in general irrational) value $E_1$ or $E_2$, and there is no 
measurement error.\footnote{
In an -- apparently nowhere explicitly discussed -- more liberal reading
of the Born rule, some additional measurement error might be acceptable.
But then Born's rule is no longer about meaurement but about idealized
measurements, whose observations are theoretical numbers, not 
actual results. Thus the liberal reading of Born's rule would be a
purely theoretical construct, silent about actual measurement results.
} 
However, the measurement result is not reproducible: Multiple 
repetition of the measurement results in a random sequence of values 
$E_1$ and $E_2$,with probabilities $p$ and $1-p$, respectively. In the 
limit of arbitrarily many repetitions, the mean value of this sequence 
approaches $\ol A$ and the standard deviation approaches $\sigma_E$.

If the energy levels are exactly known beforehand (or if the ''energy'' 
actually represents a component of a spin variable), one can calibrate 
the pointer scale to make  $E_1'=E_1$ and $E_2'=E_2$. Then, as long as 
one ignores the idealization error, both interpretations become 
experimentally indistinguishable. 
However, as already pointed out in Part I \cite{Neu.Ifound}, in the 
more realistic case where energy levels are only approximately known 
and must be inferred experimentally -- the common situation in 
spectroscopy. The thermal interpretation, in agreement with the 
standard recipes for drawing inferences from inaccurate measurement 
results, still gives a correct account of the actual experimental 
situation, while Born's statistical interpretation paints an inadequate,
idealized picture only.

\subsection{The double slit  experiment}\label{ss.doubleSlit}

Consider the quantum system consisting of the screen and an external 
classical electromagnetic field. This is a very good approximation to 
many experiments, in particular to those where the light is coherent. 
According to the standard interpretation, the analysis (given, e.g., in 
the quantum optics book by \sca{Mandel \& Wolf} \cite[Chapter 9]{ManW})
of the response of the electrons in the screen to the field gives a 
Poisson process for the electron emission, at a rate proportional to 
the intensity of the incident field. This is consistent with what is 
observed when doing the experiment with coherent light. A local 
measurement of the parameters of the Poisson process therefore provides 
a measurement of the intensity of the field.

In tis analysis, there is nothing probabilistic or discrete about the
 field; it is just a term in the Hamiltonian of the system. Thus, 
according to the standard interpretation, the probabilistic response 
is in this case solely due to the measurement apparatus -- the screen, 
the only quantum system figuring in the analysis. At very low intensity,
the electron emission pattern appears event by event, and the 
interference pattern emerges only gradually. Effectively, the screen 
exhibits what is called \bfi{shot noise}: it begins to stutter like a 
motor when fed with gas at an insufficient rate. The stuttering of the 
screen cannot be due to discrete eigenvalues of an operator representing
the intensity -- the only operator appearing in the analysis by Mandel 
and Wolf is an electron momentum operator coupling to a classical field.

The classical external field discussed so far is of course only an 
approximation to the quantum electromagnetic field, and was only used 
to show that the discrete response is due to the detector, and only 
triggered by the interaction with a field. A field mediating the 
interaction must be present with sufficient intensity to transmit the 
energy necessary for the detection events; these are for coherent 
quantum light independent and Poisson distributed even in a full quantum
analysis (given by \sca{Mandel \& Wolf} \cite[Section 12.10]{ManW}).
In the case of noncoherent quantum light, only the quantitiative 
details change. 

The discrete result appears just because each screen electron makes a 
very inaccurate random binary measurement of the incident field 
intensity. Each single spot in the gradually appearing interference 
pattern is measurable to high accuracy, but this is a high accuracy 
measurement of the screen only, not of the field (or its particle 
content). The low accuracies refer to accuracies of the implied field 
intensity -- namely one unit at the responding position and zero units 
elsewhere, while the true intensity is low but nonzero everywhere where 
the high intensity interference pattern would show up.

Accepting Mandel and Wolf's detector analysis, nothing depends on the 
deterministic nature of the thermal interpretation. But the latter 
explains (see Subsection \ref{ss.emergentBorn}) why neglecting the 
environment results in probabilistic features at all, and causes the 
electrons to exhibit a binary response -- remaining bound or escaping 
to a macroscopic distance where the effect can be magnified by a 
photomultiplier.

\subsection{Quantum buckets}\label{ss.bucket}

In the thermal interpretation, one assumes that a stuttering effect 
similar to the one discussed in the preceding subsection, when measuring
a low intensity classical electromagnetic field by a photosensitive 
surface, appears whenever one measures any classical or quantum field 
at very low intensity, whether a photon field or an electron field or 
a silver field or a water field is considered. 

The stuttering effect mentioned in Subsection \ref{ss.doubleSlit} 
may be illustrated as follows. We consider measuring 
the rate of classical water flow into a basin by the number of buckets 
(of a fixed size) per unit time needed to keep the water at a roughly 
fixed level of height. As long as there is enough flow the bucket is 
very busy and the flow is measured fairly accurately. But at very low 
rates it is enough to occasionally take out one bucket full of water 
and the bucket number is a poor approximation of the flow rate unless 
one takes very long times.

By the same principle, quantum detectors such as photocells and Geiger 
counters act as \bfi{quantum buckets}. The sole fact that one has 
counters already implies that, whatever they measure, the measurements 
are forced by construction to be integers. This limits the attainable 
resolution of what is measured as in the example of the 3-digit counter 
from Subsection \ref{ss.true}. If used to measure continuous flow, the 
uncertainty is always at least $1/2$ in the units used for the counting.

\section{Currents and particles}\label{s.currents}

In this section we look at how situations traditionally treated in 
terms of currents may be viewed in the thermal interpretation in terms 
of fields or, more precisely, currents. 

After cosidering the notion of currents in general, we look at how they 
may be used to visualize particle decays. 

We then consider the Stern--Gerlach experiment, one of the standard 
textbook examples used in the context of introducing Born's rule. 
Here a silver beam is split by a magnetic field into two beams. These 
beams are observed to produce two spots of silver deposit on a screen. 
With the thermal interpretation, we may interpret this experiment 
either on the level of quantum field theory in terms of currents or by 
considering individual silver atoms in the beam. The former is the 
fundamental level and is treated in Subsection \ref{ss.SGcurrents}. 
The latter is approximate but elementary and is treated in Subsection 
\ref{ss.SGparticles}.

\subsection{Currents}\label{ss.currents}

We consider the example for the measurement of a current with a 
galvanometer. From a quantum field theoretical point of view, an 
electric current consists (in the situation to be discussed here) of 
the motion of the electron field in a wire at room temperature. 
The thermal interpretation says that at any level of description, one 
has an electron field, and the theoretically exact current density is 
described by the distribution-valued beable (q-expectation)
\[
J^{\mu}(x) = \Tr\rho j^{\mu}(x)
\]
determined by the current operator 
$j^{\mu}(x)=-e :\bar{\psi}(x) \gamma^{\mu} \psi(x):$. Here the colons 
denote normal ordering and $\rho$ is the density operator describing 
(in the Heisenberg picture) the exact state of the universe. Denoting by
$\kbar$ the \bfi{Boltzmann constant}, we define the 
\bfi{entropy operator} of the universe by $S:=-\kbar\log\rho$, so that
$\rho=e^{-S/\kbar}$. 

At this level of description there is no approximation at all; the 
latter is introduced only when one replaces the exact $S$ by a 
numerically tractable approximation. At or close to thermal equilibrium,
it is well-established empirical knowledge that we have 
\[
S \approx (H+PV-\mu N)/T;
\]
equality defines exact equilibrium. We can substitute this (or a more 
accurate nonequilibrium) approximation into the defining formula for 
$J(x)$ to compute a numerical approximation.

Ignoring reading uncertainties, a galvanometer measures an electric
current of the form 
\lbeq{e.effCurrent}
I(t) =\int dz\ h_t(z)\cdot J(x+z)
\eeq
flowing at time $t$ through a cross section of the galvanometer. 
Here $h_t(z)$ is a smearing function that is negligible for $z$ larger 
than the size of the current-sensitive part of the galvanometer. 
The precise $h$ can be found by calibration.

The smearing is needed for mathematical reasons to turn the 
distribution-valued current into an observable vector, and for physical 
reasons since the galvanometer is insensitive to very high spatial or 
temporal frequencies. This smearing has nothing to do with 
coarse-graining: It is also needed in already coarse-grained classical 
field theories. For example, in hydromechanics, the Navier-Stokes 
equations generally have only weak (distributional) solutions that 
make numerical sense only after smearing.

Thus the quantum situation is in the thermal interpretation not very 
different from the classical situation. In particular, nowhere was made 
use of any statistical argument; the trace (which in traditional 
statistical mechanics gets a statistical interpretation) is simply a 
calculational device for managing the q-expectations.

\subsection{Particle decay} 

For other currents everything is analogous. It will be shown elsewhere
\at{cf. Haag and Ruelle,  Sandhas} 
that one can canonically associate to every bound state of a 
Poincar\'e invariant relativistic or Galilei invariant nonrelativistic
quantum field theory a distinguished effective 4-vector current 
operator. This allows one to represent all asymptotic scattering 
phenomena at finite times using currents in place of particles. 

In particular, in the thermal interpretation, currents provide the 
natural description for chemical reactions, collision processes, and 
particle decay, using the general picture justified in Subsections 4.2 
and 5.1 of Part III \cite{Neu.IIIfound} that discrete events emerge from
coarse-graining through dissipation together with the discrete basin 
structure of the slow manifold of a physical system. 

We explain the principle by considering a particle decay $A\to B+C$, 
such as $\pi^{+} \rightarrow \mu^{+} + \nu_{\mu}$.
Note that at present, this only gives an intuitive picture of what 
should happen. The details of this thermal interpretation picture are 
still conjectural and need to be justified by future analysis of 
specific models. 

At each time $t$ one has three operator-valued effective 4-currents, 
one for each possibly flowing substance $A,B,C$. When the center of the 
reaction is at the origin, the reaction $A\to B+C$ proceeds as follows: 
At large negative times the $A$-density 
(q-expectation of the time component of the 4-current) is concentrated 
along the negative z-axis, and the $A$-current (q-expectation of the 
3-vector of space components of the 4-current) is concentrated along 
the positive z-axis; the $B$-current and the $C$-current essentially 
vanish.

If the reaction happened (which depends on the details of the 
environment) then, at large positive times, the $A$-current is 
negligible, the $B$-density and $C$-density are concentrated along two 
(slightly diverging) rays emanating from the origin in such a way that 
momentum conservation holds, and the $B$-current and $C$-current are 
concentrated along these rays, too. Otherwise, at large positive times, 
the $A$-density is concentrated along the positive z-axis, and the 
$A$-current is concentrated along the positive z-axis, too, and the 
$B$-current and the $C$-current remain negligible. During the reaction 
time, i.e., when the fields are concentrated near the origin, one can 
interpolate the asymptotic happening in an appropriate way. 
The details are defined by the interaction.

The manifold of slow modes splits into a basin corresponding to the 
decayed state (with two continuous angle parameters labeling the 
possible modes) one basin corresponding to the undecayed state. 
The metastable transition state at time zero determines together with 
the environmental fluctuations which basin is chosen and which direction
is taken. This is comparable to what happens to bending a classical 
thin iron bar through longitudinal pressure in a random direction, 
though in that case the bar must bend, so that there is only one basin, 
with modes labelled by a single angle. In both cases, one of the 
continuous labels appears due to the rotational symmetry of the setting 
around the z-axis. In the case of the decay reaction, the second 
continuous label arises through another, infinitesimal symmetry at the 
saddle point at the origin.

This is one of the possible scenarios, probably what happens if the 
decay happens inside a dense medium (a secondary decay in a bubble 
chamber, say).

For a collision experiment in vacuum, there is probably not enough 
environmental interaction near zero, and after reaching the collision 
region, the $B$-current and the $C$-current should, in case a reaction 
happens, rather take a rotationally symmetric shape. In this case, the 
path like particle nature appears only later when the spherical fields 
reach a detector. The metastability of the detector forces the two 
spherical fields to concentrate along two paths, and momentum 
conservation makes these paths lie weighted-symmetric to the z-axis 
(would be geometrically symmetric when the decay products have equal 
mass). The details are essentially those reported in the 1929 paper by 
\sca{Mott} \cite{Mott}.

In both cases, the detection process creates the seeming particle 
nature of the observation record; cf. the discussion in Subsection 4.4
 of Part I \cite{Neu.Ifound}.

\subsection{The Stern--Gerlach experiment in terms of currents}
\label{ss.SGcurrents} 

In the traditional analysis of the Stern--Gerlach experiment in terms 
of single silver particles, the dynamics is treated semiclassically 
for simplicity, and two beams appear as the only possible pathways. 

In a field theoretic treatment, the beam is not interpreted classically 
but as a quantum field.\footnote{
This is the difference to Schr\"odinger's failed early attempts to give 
a continuum interpretation of quantum mechanics in terms of classical 
fields.
} 
Thus the silver is treated as an effective spinor field. It is not a 
free field because of the magnetic field in the experiment. The 
magnetic field is (in the usual semiclassical treatment) a term in the 
Hamiltonian of the field theory that changes the dynamics. It  treats 
different components of the spinor field representing silver in 
opposite ways, turning a single beam at the source into two while 
passing the magnet. 

Subsection \ref{ss.currents} applies, except that the electric current 
is replaced by a silver current (which means that the formula defining 
it is a complicated multibody current). To get the total amount of 
silver deposited one also  needs to integrate over the time of the 
experiment. Thus the effective support of the silver current operator 
$j(x)$ is initially along a single beam, which, upon entering the 
magnetic field, splits into two beams. The current flows along the 
direction of the two beams.  The amount of silver on the screen at the 
end measures the integrated beam intensity, the total transported mass.
This is in complete analogy to the qubit treated in Subsection 3.5 of
Part III \cite{Neu.IIIfound}. Particles need not be invoked.

The intensity of silver flow is the function of the position on the 
screen defined by the q-expectation of the incident current integrated 
over a spot centered at this position. Given the setup, the intensity 
is positive at the two spots predicted by the mathematics of the theory,
and zero elsewhere. 

The density operator is that of the whole universe, and the integration 
in \gzit{e.effCurrent} is effectively over a cell to which a piece of 
the equipment responds, done after the trace computation. The operation 
$\Tr\rho j(x)$ yields a current $J$ that is nonzero only at two small 
spots of any cross section (e.g., on the screen), and integrating over 
each spot gives in the symmetric case a total intensity of half of the 
original beam (before the apparatus) in each spot. Integrating over 
other regions of the screen gives zero since the integrand is zero 
there. This is why the silver flows into these two spots and nowhere 
else. 

Thus when firing a continuous beam of high intensity one sees two spots,
both appearing at essentially the same time. 
What is measured by a spot is the intensity of the silver flow 
into the spot, not the spin of single electrons.\footnote{
The original Stern--Gerlach paper (and the early discussion about it) 
indeed talked about ''Richtungsquantelung'' (quantization of directions)
and not of spin measurement. The fact that two beams appear is a 
consequence of the spin of the electron field, but has nothing per se 
to do with measuring an electronic spin state. The latter is defined 
only for single electrons, not for the electron field.
} 

In the very low density case, the stuttering effect discussed in 
Subsection \ref{ss.doubleSlit} for the double slit experiment becomes 
visible at a screen of sufficiently high resolution, and the response
of the screen becomes erratic. In particular, if a beam contains only a 
single particle, the quantum field representing the beam is in a state 
with sharp particle number $N=1$, but otherwise nothing changes. 
Conservation of mass, together with the instability of macroscopic 
superpositions and randomly broken symmetry forces that only one of 
the two spots gets marked by a silver atom, just as a classical bar 
under vertical pressure will bend into only one direction. It is not
clear how Nature achieves the former, but this lack of explanation is 
common to all interpretations of quantum physics. 

We may interpret the stuttering in terms of the quantum bucket picture 
from Subsection \ref{ss.bucket}. We may think of each of the two spots 
on the screen as a quantum bucket counting impinging silver flow. 
We combine the counts into a single pointer variable $x$ by counting 
left spot events downwards ($-1$) and right spot events upwards ($+1$). 
Each single atom deposited somewhere on the screen is one bucket event 
reducing the intensity of the inflowing bilocal silver field. It 
approximates the true value in $[-1,1]$ (the q-expectation of $x$) by 
either $+1$ or $-1$, the only possible bucket results. This holds for 
every single atom, and hence for all the silver that arrives in the two 
spots. If we assume for simplicity that the silver source is prepared 
in a state where the q-expectation of $x$ vanishes, taking the single 
buckets as measurement results each time in a binary measurement of the 
true (theoretically predicted) uncertain number $0\pm 1$, consistent 
with the measurement error of 1 in each case. This is completely 
independent of the flow rate.

\subsection{The Stern--Gerlach experiment in terms of particles}
\label{ss.SGparticles}

\nopagebreak
\hfill\parbox[t]{10.8cm}{\footnotesize

{\em An dem Tatbestand, die Elektronenschw\"arme betreffend, wie er
bisher beschrieben wurde, ist nichts Paradoxes. Statt vom Schwarm
spreche ich in Zukunft vom einzelnen Elektron und demgem\"a{\ss} von
Wahrscheinlichkeit statt von H\"aufigkeit. Etwas Paradoxes liegt erst 
in der Aussage, da{\ss} $\sigma_x$ die Komponente eines gewlssen 
Vektors, des Impulsmomentes, in bezug auf die $x$-Richtung ist. Denn 
dies involviert doch, wenn wir eln rechtwlnkliges Koordinatensystem 
$x\,y\,z$ im Raume einf\"uhren und die willk\"urliche Richtung $r$ die 
Richtungskosinus $a, b, c$ hat, die Gleichung
\[
\sigma_r =a \sigma_x +b \sigma_y + c \sigma_z.
\]
Wie vertr\"agt sich das mlt dem Umstand, da{\ss} $\sigma_r$ so gut wie 
$\sigma_x, \sigma_y, \sigma_z$ nur der Werte $\pm 1$ f\"ahlg ist?
}

\hfill Hermann Weyl, 1927 \cite[p.8f]{Weyl1927}
}
\bigskip

We now consider the Stern--Gerlach experiment not in terms of a field 
measurement but as a spin measurement experiment of single silver atoms 
in the beam. In this case one must -- like in every introductory text --
treat the silver source as producing an ensemble of single atoms and, 
ignoring efficiency considerations, assume that each silver atom 
produces a tiny dot at one of the two spots on the screen. 

The thermal interpretation looks at the reduced dynamics of the 
relevant macroscopic q-expectations and would find by a similar 
analysis as that in Section \ref{s.singleQubits} that due to the reduced
dynamics of the pointer variable (here the relative position of the 
condensed silver atom), all but the positions at the two spots are 
unstable, so that the total system is bistable. 

Ignoring a factor of $\hbar/2$, we represent the spin measurement as 
a measurement of the q-expectations $\<\sigma_3\> \in[-1,1]$ by means
of a binary measurement of the spot on which an arriving silver atom is 
located, with possible values left spot ($-1$) or right spot ($+1$).
In the thermal interpretation, each single dot on the screen, at either 
the left spot ($-1$) or the right spot ($+1$), is viewed as an 
approximate 
measurement of the q-expectation, which lies somewhere in $[-1,1]$.
This approximation is very poor. For example, when the initial state of 
the silver atoms is such that its q-expectations is $\<\sigma_3\>=0$, 
the error of both binary measurement results $\pm 1$ is $1$, but with 
random signs, consistent with the computed uncertainty, which is also 
$1$.

To improve the accuracy one needs to average over multiple measurement, 
and gets better results that converge to the true value 0 as the sample 
size gets arbitrarily large. To see this, one must consider a different 
operator, namely the mean spin $s=N^{-1}(s_1+\ldots+s_N)$, where $s_k$ 
is the $\sigma_3$ of the $k$th silver atom in the ensemble measured. 
This mean spin operator has an associated (theoretically predicted) 
uncertain value of $\ol s\pm \sigma_s=0\pm N^{-1/2}$, which is 
approximately measured by 
the mean of the bucket results. This mean is for large $N$ distributed 
as a Gaussian with zero mean and standard deviation $N^{-1/2}$, matching
the prediction.

\bigskip

Born's statistical interpretation treats the measured position of each 
individual silver atom instead as an exact measurement of the discrete 
value $\pm 1$ of the corresponding atom, with random signs. Although 
each single measurement is deemed error-free, the statistical 
uncertainty resulting from this randomness is still $1$.

Clearly, both interpretations account for the same experimental facts, 
but in different ways. They make very different assumptions concerning 
the nature of what is to be regarded as the idealized measurement result
to which the actual result is to be compared.

\bigskip

\end{document}